\documentclass[apj]{emulateapj}\usepackage{apjfonts}

\newcommand{\kms}{\ifmmode {\rm km\,s}^{-1} \else km\,s$^{-1}$\fi}

\shorttitle{OH 1720\,MHz masers in Sgr\,A\,East and the CND} 
\shortauthors{Sjouwerman \& Pihlstr\"om}

\begin{document}

\title{Very Large Array Observations of Galactic Center OH 1720\,MHz Masers
  in Sagittarius\,A\,East and in the Circumnuclear Disk}

\author{Lor\'{a}nt~O.~Sjouwerman}
\affil{National Radio Astronomy Observatory, 1003 Lopezville Rd.,
  Socorro, NM 87801,} 
\author{Ylva~M.~Pihlstr\"{o}m}
\affil{Department of Physics and Astronomy, University of New Mexico,
  800 Yale Boulevard NE, Albuquerque, NM 87131,} 
\email{lsjouwer@nrao.edu, ylva@unm.edu.}

\begin{abstract}

We present Very Large Array (VLA) radio interferometry observations of
the 1720\,MHz OH masers in the Galactic Center (GC). Most 1720\,MHz OH
masers arise in regions where the supernova remnant Sgr\,A\,East is
interacting with the interstellar medium. The majority of the newly
found 1720\,MHz OH masers are located to the northeast, independently
indicating and confirming an area of shock interaction with the $+$50
\kms\ molecular cloud (M$-$0.02$-$0.07) on the \emph{far} side of
Sgr\,A\,East. The previously known bright masers in the southeast are
suggested to be the result of the interaction between two supernova
remnants, instead of between Sgr\,A\,East and the surrounding
molecular clouds as generally found elsewhere in the Galaxy. Together
with masers north of the circumnuclear disk (CND) they outline an
interaction on the \emph{near} side of Sgr\,A\,East.  In contrast to
the interaction between the $+$50 \kms\ cloud and Sgr\,A\,East, OH
absorption data do not support a direct interaction between the CND
material and Sgr\,A\,East.  We also present three new high-negative
velocity masers, supporting a previous single detection. The location
and velocities of the high-negative and high-positive velocity masers
are consistent with being near the tangent points of, and physically
located \emph{in} the CND.  We argue that the high velocity masers in
the CND are pumped by dissipation between density clumps in the CND
instead of a shock generated by the supernova remnant. That is, the
CND masers are not coupled to the supernova remnant and are sustained
independently.
\end{abstract}


\section{Introduction}\label{intro}

Two different types of masers are observed in the 1720\,MHz satellite
line transition of hydroxyl (OH); one is found in star forming regions
(SFRs) and the other is associated with supernova remnants (SNRs). The
radiatively pumped 1720\,MHz masers in SFRs are accompanied by masers
in the other ground-state rotational transitions (at 1612, 1665 and
1667\,MHz) as a result of the cascade down from higher excitation
levels, whereas the collisionally pumped 1720\,MHz line is the only OH
transition observed near SNRs \citep{lockett99, wardle99,
pihlstrom08}. The latter masers originate in post-shocked regions
where an expanding SNR collides with dense molecular clouds in the
surrounding interstellar medium (ISM), and are observed near SNR/ISM
interaction regions throughout the Galaxy \citep[e.g.,][]{frail94,
green97}. The Galactic center (GC) with the Sgr\,A\,East SNR plowing
into the ISM surrounding the Sgr\,A\,Complex is no exception.


\begin{table*}
\footnotesize
\begin{center}
\caption{The observing dates, VLA baseline configuration, channel
separation $\Delta V$, total velocity range observed, typical
1$\sigma$ rms noise level per channel, synthesized beam size plus beam
position angle for the four VLA data sets and remarks on each data
set.}
\smallskip
\begin{tabular}{lcccccl} \hline \hline
Date of observation & Config. & $\Delta V$ & $V$-range
& Channel rms & Beam size \& P.A. & Remarks\\
& &  {\small(\kms)} & {\small(\kms)} & {\small (mJy beam$^{-1}$)} &
{\small (\arcsec\ $\times$\arcsec\ at $^\circ$)} &\\
\hline
1986 Jun 30&BnA&8.5&$-$276 to $+$200&$\phd$4&$\phantom{0}$5.2$\times$4.3, $\phantom{-}$10.4&\citet{karlsson03}\\
1996 Nov 22&BnA&2.1&$-$194 to $+$194&19&$\phantom{0}$7.1$\times$1.3, $\phantom{-}$39.8&Unpublished$^a$ part of \citet{yusef-zadeh99}\\
1998 Jul 06&B  &8.5&$-$276 to $+$200&$\phd$3&10.4 $\times$4.4, $\phantom{-}$23.3&Previously unpublished archival data\\
2005 Jan 20 \& 26&BnA&2.1&$-$232 to $+$232&$\phd$3&$\phantom{0}$3.9 $\times$ 3.6, $-$57.9&New observations\\
\hline 
\multicolumn{7}{l}{$^a$ \citet{yusef-zadeh99} used two observational set-ups, only the 
                   narrow bandwidth high spectral resolution data were published in their paper.}\\
\end{tabular}\label{datasets}
\end{center} 
\end{table*}


The line-of-sight toward the Sgr\,A\,Complex consists of the SNR
Sgr\,A\,East and a circumnuclear disk (CND). Sgr\,A\,East manifests
itself as a radio continuum ridge or shell \citep[e.g.,][]{ekers83,
nord04}, in part obscured by the torus- or ring-like CND. The CND
consists of irregularly distributed clumps of molecular gas
\citep[e.g.,][]{jackson93,christopher05} rotating counter-clockwise in
a ring with a mean radius of about 2 pc around the compact radio
source Sgr\,A*, the dynamical center of the Milky Way
\citep{reid04}. Interior to the CND, most gas is ionized and is
distributed in a ``minispiral'', also known as the \ion{H}{2} region
Sgr\,A\,West.  This line-of-sight also partly overlaps with two Giant
Molecular Clouds (GMCs) casually called the $+$20 and $+$50 \kms\
clouds \citep[M$-$0.13$-$0.08 and M$-$0.02$-$0.07; see reviews
in][]{morris96, mezger96}. These form the ``molecular belt''
stretching across the Sgr\,A\,Complex, providing the ISM that
interacts with Sgr\,A\,East.

It has long been known that a number of bright 1720\,MHz masers with
line-of-sight velocities near 50--65 \kms\ are observed in the
southwest region of Sgr\,A\,East, toward the SNR G359.02$-$0.09
\citep{ho85,coil00}. A few other isolated masers are also distributed
at other locations along the radio continuum shell with velocities
between 40 and 60 \kms\ (\citealt[][hereafter YZ96]{yusef-zadeh96};
\citealt[][hereafter K03]{karlsson03}). In Sect.~\ref{snrmasers} we
present new, less prominent 1720\,MHz OH masers that are due to the
interaction of Sgr\,A\,East and the $+$50 \kms\ molecular cloud
observed toward the northeast
\citep[e.g.,][]{zylka90,ho91,tsuboi06,lee08}.

In addition, a few masers with very high absolute line-of-sight
velocities can be found, at least in projection, near the CND (YZ96;
K03; this work). Their high velocities (about $+$130 and $-$130 \kms)
do not fit the SNR/ISM interaction model.  Instead, their signature
resembles the structures of 1667\,MHz OH and 22\,GHz H$_2$O masers
found in circumnuclear tori of nearby galaxies \citep{miyoshi95,
pihlstrom01,kloeckner03} suggesting that these masers arise in the CND
itself.  Until now this picture for the GC depended on a previous
single detection of a $-$132 \kms\ maser. Section~\ref{cndmasers}
presents support for this model with the detection of a number of new
high-negative velocity masers.

In Sect.~\ref{observations} and Sect.~\ref{results} we outline the
data reduction procedures and present the results. In
Sect.~\ref{discussion} the nature of the different groups of 1720\,MHz
masers in the GC is discussed. With this in mind, we present a model
of the interaction of Sgr\,A\,East with its surroundings. Unless
stated otherwise, all velocities in this paper refer to line-of-sight
velocities as measured with respect to the Local Standard of Rest
(LSR).

\section{New observations and archival data}\label{observations}

New high-sensitivity and high spatial resolution VLA observations were
obtained to search for high-velocity masers in the CND. The new, 2005
January observations (Table~\ref{datasets}) had two IF pairs of
1.562\,MHz bandwidth, each tuned to slightly offset center velocities
in order to cover a large total velocity range ($-$232 $<V_{\rm
LSR}\,(\kms) <+$232). The data were calibrated using VLARUN, a
pipeline VLA data reduction procedure available in NRAO's Astronomical
Image Processing System (AIPS). After continuum subtraction in the
UV-plane the data were imaged with natural weighting using standard
AIPS procedures, very similar to the data reduction as described in
K03. The only difference was that we joined the spectra from the two
IF pairs into one long spectrum covering the whole velocity range with
the higher spectral resolution using UJOIN. Self-calibration was
performed using the bright $+$132 \kms\ maser feature, which resulted
in positional uncertainty of about 0.2--0.3\arcsec. The resulting data
cube has a synthesized beam full width half maximum (FWHM) of 3\farcs
9 $\times$ 3\farcs 6 at a position angle (P.A.) of $-$57.9$^\circ$,
and a channel separation of 2.1 \kms\ (i.e., a channel resolution of
2.5 \kms). With dual polarizations and a total on-source integration
time of about three hours, a typical rms noise of 3.2 mJy\,beam$^{-1}$
per channel was achieved.

Three archival VLA data sets (out of which two are not published) have
comparable velocity coverage, spatial resolution and sensitivity to
the new observations and were re-analyzed following the same
calibration procedure described above (see Table~\ref{datasets}). They
were included because we were interested in searching for the newly
discovered masers in the older data sets (Sect.~\ref{resmas}), and
re-analyzed because such a search would be easier when using the same
(self-calibration) reference peak. Since the masers typically are as
narrow as a couple \kms, all masers in these archival data sets should
have been re-detected in the new data set if the masers would not vary
in flux.  All archival data sets are in B1950 coordinates, and the new
(2005 January) data set is in J2000 coordinates.


\begin{figure*} \begin{center}
\plotone{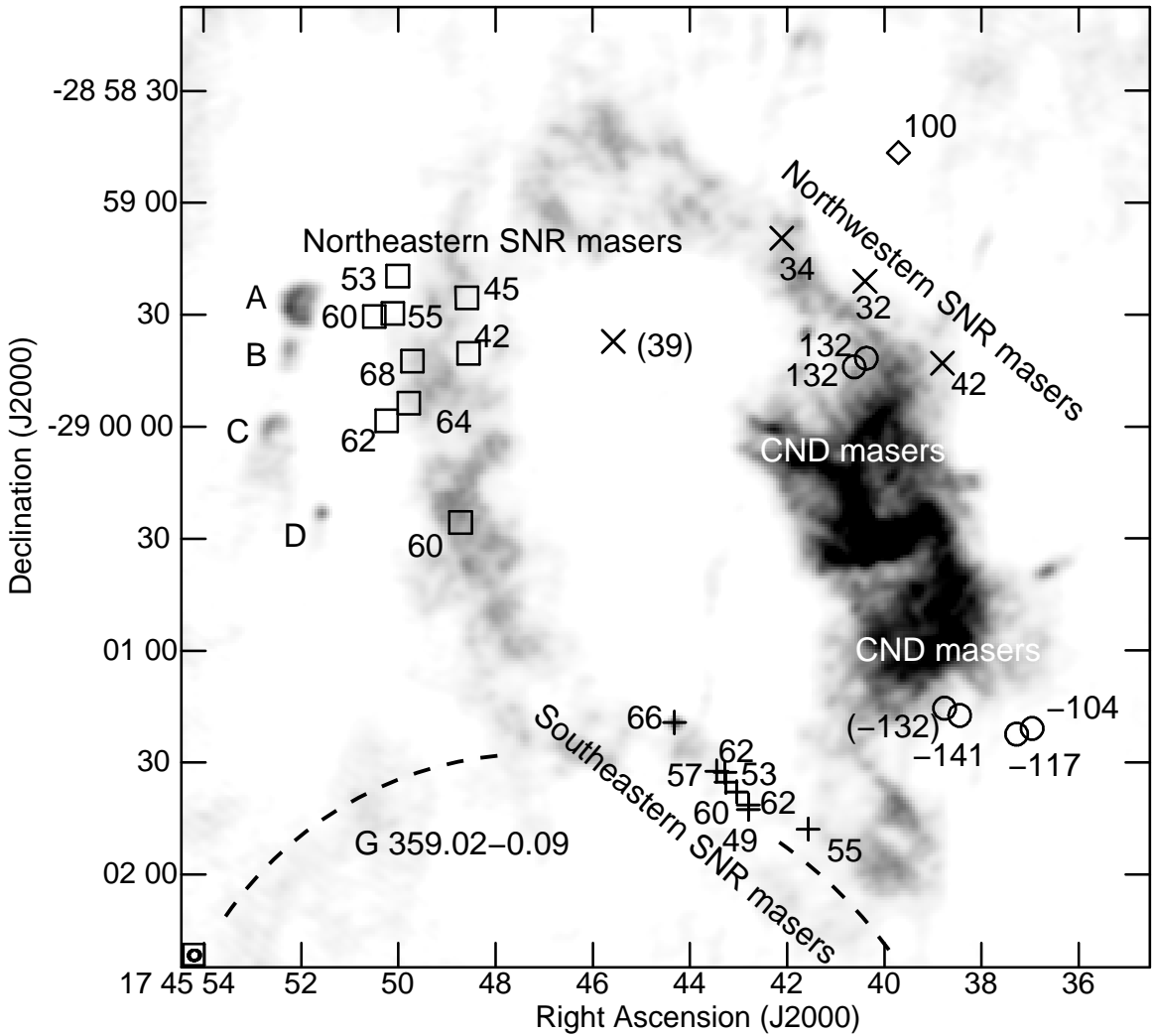}
\end{center}
\caption{Location of the 1720\,MHz OH masers with respect to
Sgr\,A\,East (ring-like structure) and the CND and minispiral (dark
S-like spiral on the right). The masers are superimposed on the 1.7
GHz radio continuum created from the frequency averaged 2005 January
data. Numbers associated with symbols are the maser LSR
velocities. Squares, crosses and plusses denote masers associated with
the SNR shell (i.e., the northeastern, northwestern and southeastern
SNR masers, respectively), circles are masers associated with the
CND. The diamond represents another newly detected maser. All masers
in the upper left (squares, northeastern SNR masers) except the $+$55
\kms\ one are newly found masers. Two masers from \citet{karlsson03}
with velocities in parentheses are not re-detected in the data after
1986.  The molecular belt roughly stretches with increasing velocity
from the lower right corner to the upper left corner.  The dotted
semi-circle outlines the SNR G359.02$-$0.09 and labels A-D identify
\ion{H}{2} regions (see Sect.~\ref{snrmasers}).
\label{maserpos} }
\end{figure*}


\begin{table*} \begin{center}
\footnotesize
\caption{1720\,MHz OH masers detected in the Sgr\,A Complex.}
\smallskip
\begin{tabular}{llrlcclcclclrlrll} \hline \hline
No & & \multicolumn{1}{c}{OH Name} & \phd &\multicolumn{2}{c}{Position in B1950} & & \multicolumn{2}{c}{Position in J2000} & 
\phd & $\Delta V^a$ & & V$_{sys}$ & & Flux & & Ref.$^b$ \\
 & &\multicolumn{1}{c}{$(l,b)$}& &\multicolumn{2}{c}{\footnotesize(RA \& Dec B1950)}& &
\multicolumn{2}{c}{\footnotesize(RA \& Dec J2000)}& &\multicolumn{3}{c}{\footnotesize(\kms)}& &{\footnotesize(Jy)} & & \\
\hline 
1      && 359.925$-$0.044 && 17\,42\,26.21 & $-$29\,00\,11.2 && 17\,45\,36.96 & $-$29\,01\,20.9 && 4.2 && $-$104 &&  0.11 &&  \\
2      && 359.926$-$0.045 && 17\,42\,26.53 & $-$29\,00\,12.7 && 17\,45\,37.28 & $-$29\,01\,22.4 && 2.1 && $-$117 &&  0.05 &&  \\
3      && 359.929$-$0.048 && 17\,42\,27.70 & $-$29\,00\,07.7 && 17\,45\,38.45 & $-$29\,01\,17.4 && 2.1 && $-$141 &&  0.04 &&  \\
4$^c$  && 359.930$-$0.048 && 17\,42\,28.02 & $-$29\,00\,05.8 && 17\,45\,38.76 & $-$29\,01\,15.5 && $\phantom{^c}$8.5$^c$ && $-$132 &&  0.07 && 1,2 \\
5      && 359.952$-$0.035 && 17\,42\,28.10 & $-$28\,58\,33.4 && 17\,45\,38.80 & $-$28\,59\,43.0 && 4.2 &&     42 &&  0.62  && 1,3 \\
6      && 359.967$-$0.030 && 17\,42\,29.02 & $-$28\,57\,37.0 && 17\,45\,39.71 & $-$28\,58\,46.6 && 2.1 &&    100 &&  0.04 &&  \\
7$^c$  && 359.955$-$0.040 && 17\,42\,29.67 & $-$28\,58\,32.2 && 17\,45\,40.37 & $-$28\,59\,41.7 && $\phantom{^c}$8.5$^c$ &&    132 &&  0.29 && 1,3 \\
8      && 359.960$-$0.037 && 17\,42\,29.71 & $-$28\,58\,11.6 && 17\,45\,40.40 & $-$28\,59\,21.1 && 4.2 &&     32 &&  0.15 && 1 \\
9$^c$  && 359.955$-$0.041 && 17\,42\,29.91 & $-$28\,58\,34.4 && 17\,45\,40.62 & $-$28\,59\,44.0 && $\phantom{^c}$8.5$^c$ &&    132 &&  0.66 && 1,3\\
10     && 359.928$-$0.062 && 17\,42\,30.81 & $-$29\,00\,38.5 && 17\,45\,41.57 & $-$29\,01\,47.9 && 6.3 &&     55 &&  0.51 && 1,3 \\
11     && 359.966$-$0.041 && 17\,42\,31.41 & $-$28\,58\,00.1 && 17\,45\,42.11 & $-$28\,59\,09.5 && 2.1 &&     34 &&  0.03 && \\
12$^d$ && 359.932$-$0.065 && 17\,42\,32.04 & $-$29\,00\,32.1 && 17\,45\,42.80 & $-$29\,01\,41.4 && 4.2 &&     62 &&  0.10 && 1,3 \\
13$^d$ && 359.931$-$0.065 && 17\,42\,32.05 & $-$29\,00\,33.2 && 17\,45\,42.80 & $-$29\,01\,42.6 && 2.1 &&     49 &&  0.09 && 1,3 \\
14$^d$ && 359.933$-$0.065 && 17\,42\,32.28 & $-$29\,00\,28.5 && 17\,45\,43.04 & $-$29\,01\,37.9 && 2.1 &&     60 &&  0.10 && 1,3 \\
15$^d$ && 359.934$-$0.065 && 17\,42\,32.51 & $-$29\,00\,26.0 && 17\,45\,43.26 & $-$29\,01\,35.3 && 4.2 &&     53 &&  0.40 && 1,3 \\
16$^d$ && 359.935$-$0.065 && 17\,42\,32.53 & $-$29\,00\,23.4 && 17\,45\,43.28 & $-$29\,01\,32.7 && 4.2 &&     62 &&  0.24 && 1,3 \\
17$^d$ && 359.935$-$0.065 && 17\,42\,32.69 & $-$29\,00\,22.9 && 17\,45\,43.45 & $-$29\,01\,32.2 && 4.2 &&     57 &&  0.60 && 1,3 \\
18     && 359.940$-$0.066 && 17\,42\,33.57 & $-$29\,00\,10.0 && 17\,45\,44.32 & $-$29\,01\,19.3 && 6.3 &&     66 &&  6.07 && 1,3 \\
19$^c$ && 359.966$-$0.055 && 17\,42\,34.88 & $-$28\,58\,28.0 && 17\,45\,45.58 & $-$28\,59\,37.2 && $\phantom{^c}$8.5$^c$ &&     39 &&  0.08 && 1  \\
20     && 359.971$-$0.065 && 17\,42\,37.84 & $-$28\,58\,31.4 && 17\,45\,48.55 & $-$28\,59\,40.4 && 2.1 &&     42 &&  0.03 &&   \\
21     && 359.975$-$0.063 && 17\,42\,37.89 & $-$28\,58\,16.6 && 17\,45\,48.59 & $-$28\,59\,25.5 && 4.2 &&     45 &&  0.10 &&   \\
22     && 359.961$-$0.072 && 17\,42\,37.99 & $-$28\,59\,16.8 && 17\,45\,48.72 & $-$29\,00\,25.7 && 2.1 &&     60 &&  0.03  &&  \\
23     && 359.973$-$0.069 && 17\,42\,39.00 & $-$28\,58\,33.6 && 17\,45\,49.71 & $-$28\,59\,42.4 && 4.2 &&     68 &&  0.18  &&  \\
24     && 359.970$-$0.071 && 17\,42\,39.07 & $-$28\,58\,44.8 && 17\,45\,49.78 & $-$28\,59\,53.7 && 2.1 &&     64 &&  0.08 &&  \\
25     && 359.979$-$0.067 && 17\,42\,39.31 & $-$28\,58\,10.8 && 17\,45\,50.01 & $-$28\,59\,19.6 && 2.1 &&     53 &&  0.06 &&   \\
26     && 359.977$-$0.068 && 17\,42\,39.41 & $-$28\,58\,20.8 && 17\,45\,50.11 & $-$28\,59\,29.7 && 2.1 &&     55 &&  0.04 && 1 \\
27     && 359.970$-$0.073 && 17\,42\,39.52 & $-$28\,58\,49.6 && 17\,45\,50.24 & $-$28\,59\,58.4 && 2.1 &&     62 &&  0.06 &&  \\
28     && 359.977$-$0.070 && 17\,42\,39.78 & $-$28\,58\,21.7 && 17\,45\,50.49 & $-$28\,59\,30.5 && 2.1 &&     60 &&  0.05 &&  \\
\hline
\multicolumn{17}{l}{$^a$ The velocity width is the total velocity range of the channels in which maser emission occurs.}\\
\multicolumn{17}{l}{$^b$ References: 1) \citet{karlsson03}, 2) \citet{yusef-zadeh01} (using the same data as \citealt{karlsson03}), 
                                     3) \citet{yusef-zadeh96}.}\\
\multicolumn{17}{l}{$^c$ Parameters adopted from \citet{karlsson03}; the $\Delta V$ of 8.5 \kms\ is an upper limit due to the 
                                     large channel separation used.}\\
\multicolumn{17}{l}{$^d$ Maser complex \#12--17 corresponds to maser 7 in \citet{karlsson03} and D, E, and F in \citet{yusef-zadeh96}.}\\
\end{tabular} \label{masers}
\end{center}
\end{table*}


\section{Results}\label{results}

The observational results can be divided in two parts: point-like
1720\,MHz (maser) emission and extended 1720\,MHz absorption,
presented in turn below.

\subsection{1720\,MHz masers}\label{resmas}

To unambiguously exclude noise spikes as false maser detections, we
consider a maser to be detected only if it has a channel flux density
larger than 10 times the 1$\sigma$ rms noise level in its data
set. With this strict constraint, we detect 13 new and 13 previously
published masers in the 2005 data. One of the newly discovered masers
was also detected in the 1998 data (maser \#1 in Table~\ref{masers} at
11$\sigma$) but no other new masers were found above 10$\sigma$ in the
archival VLA data. Apart from the general possibility of intrinsic
variability (which exists: Sect.~\ref{cndmasers}), this is most likely
due to the much larger channel widths (8.5 \kms) or the much larger
rms noise in the archival data sets. For example, assuming no
variability, the brightest new single-channel detection (maser \#24 at
25$\sigma$) would be smeared to less than 7$\sigma$ rms in a 8.5 \kms\
channel in the 1986 June data set and thus would remain undetected
using our strict 10$\sigma$ constraint. Actually, knowing the
positions and velocities of the newly detected masers, we found most
of the masers \#20--28 at about 4$\sigma$--6$\sigma$ levels back in
both the 1986 and 1998 data; the 1996 data was too noisy. Also maser
\#6 was found in the archival data, but at 7$\sigma$ in the 1998 data
only; it had not appeared in 1986 and apparently weakened since
1998. Two previous detections (masers \#4 and \#19 in
Table~\ref{masers}) were not detected (down to 4$\sigma$) in the 2005
data but are included in Table~\ref{masers} for the completeness of
the discussion in this paper.

Table~\ref{masers} lists the OH name in galactic coordinates, the
position of the maser, the total width in \kms\ of the channels in
which emission occurs, the velocity of the channel containing the peak
flux, the total detected flux integrated over the velocity width, and
finally references to the papers where previous detections are
reported. For the masers detected multiple times, we have adopted the
parameters derived from the most recent 2005 dataset, or from K03 if
the maser was confused or undetected in 2005 \citep[for confusion
see][]{yusef-zadeh99}. Small positional offsets compared to earlier
papers are due to differences in the self-calibration.  Velocity
offsets and flux differences are due to different observing parameters
per data set and possible intrinsic variability of the maser.  Because
the maser emission is generally found only in 1-2 channels and may be
non-Gaussian and narrower than the channel separation, spectral
fitting to the feature to obtain a fitted flux, center velocity, FWHM
and their associated errors is meaningless. We therefore assume an
uncertainty in the center velocity and velocity range of half the
channel separation (i.e., 1 \kms\ except for the ones adopted from K03
where it is about 4 \kms), and up to 50\% uncertainty in the flux. The
flux values given should therefore be regarded as indicative for their
relative intensity only. Good fits and associated errors can be
obtained with new high spectral resolution observations, e.g. as in
\citet{yusef-zadeh99}, but this is beyond the scope of this paper.

Figure \ref{maserpos} shows the masers from Table~\ref{masers} labeled
with their velocities on top of a gray scale 1.7\,GHz radio continuum
image made from the frequency averaged 2005 data. Two different
populations of masers can be identified: a) masers with velocities
between $+$30 and $+$70 \kms\ following the outline of the
Sgr\,A\,East SNR shell (``SNR masers'', marked with squares, crosses
and pluses), and b) masers with absolute velocities $>$ 100 \kms\
which are consistent with velocities of the CND at their projected
position (``CND masers'', circles). Note that the latter are located
in projection, not only on top and near the CND, but also on top and
near the Sgr\,A\,East SNR shell, albeit with a distinct high
velocity. One additional new maser (at $+$100 \kms) seems distinct
from the two main populations (diamond, Sect.~\ref{snrnw}).

We find a significant number (3) of new high-negative velocity masers
toward the southwestern part of the CND (circles in Fig.~1), but no
new high-positive velocity masers. The CND masers are only found near
the tangent points\footnote{Assuming an inclined circular torus, the
tangent points are the narrowest part at the ends of the major axis of
the projected ellipse on the sky. These are also the regions where the
projected column densities, and thus also the path-lengths for maser
gains, are highest \citep{parra05}, and where limb-brightening of a
homogeneously emitting torus occurs \citep{yusef-zadeh01}.} of the
torus-like and inclined CND, suggesting that the masers are found only
there due to the longer path-length \citep[][see
Sect.~\ref{pump}]{pihlstrom01, parra05}. Furthermore, almost all new
SNR masers are found toward the northeastern rim of the SNR shell, in
projection close to the \ion{H}{2} regions A--D (that lie in front of
the 50 \kms\ cloud), and close to where the deepest 1667 MHz OH
absorption occurs (K03). As these masers are much weaker than the
masers toward G359.02$-$0.09 in the southeast (e.g., YZ96), they have
not been detected in previous observations other than a single one in
K03.  It is in this region that \citet{ho91} and, more recently,
\citet{lee03} and \citet{tsuboi06} have detected signs of shock
interactions using observations of NH$_3$, H$_2$ and CS emission
(Sect.~\ref{snrmasers}).

\subsection{1720\,MHz OH distribution}\label{resabs}

Although individual clumps and streamers do confuse and complicate the
overall interpretation, our 1720\,MHz OH absorption maps are
completely consistent with the conclusions drawn by K03 and others
\citep[e.g.][]{pedlar89,sandqvist89,zylka90}. That is, \emph{i)} part
of the distribution of the $+$20 and $+$50 \kms\ clouds is located
\emph{in front of} the Sgr\,A\,East SNR shell, \emph{ii)} the CND must
be located on the near side of Sgr\,A\,East, and, \emph{iii)} the
$+$20 \kms\ and $+$50 \kms\ clouds must be {\em mostly behind} Sgr\,A*
and the minispiral, thus behind the CND.

\begin{figure*} \begin{center}
\plotone{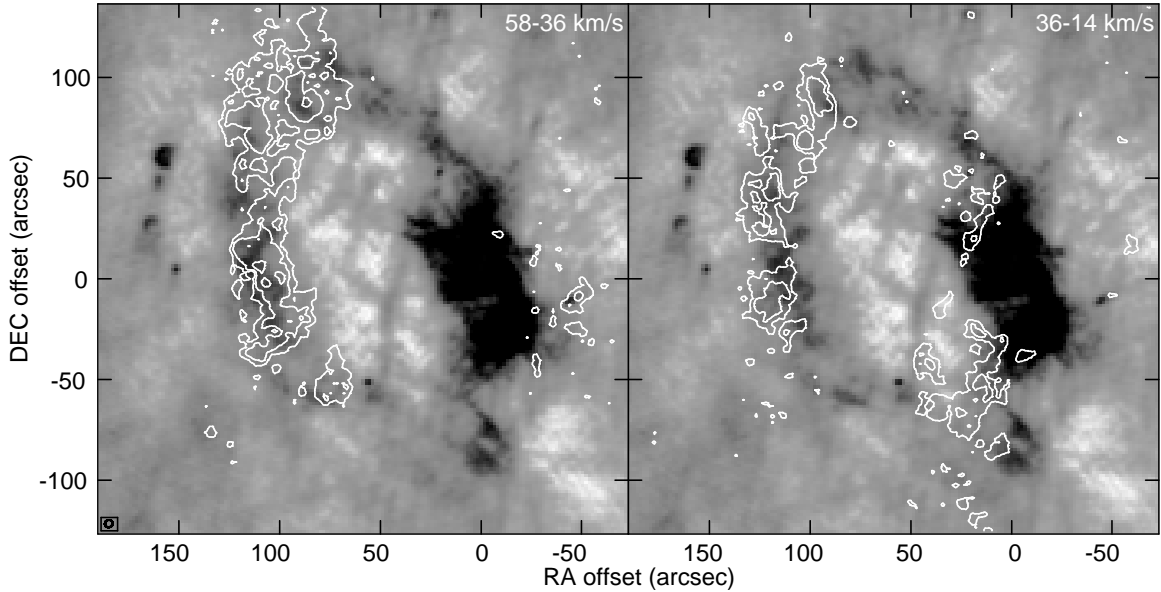}
\end{center}
\caption{Spectrally smoothed 1720\,MHz OH absorption at velocities
between 58-36 and 36-14 \kms. The data was smoothed to improve the
signal to noise ratio. The absorption is shown in contours at 7.5, 15
and 22.5 mJy\,beam$^{-1}$, superimposed on the gray scale image of a
1.7\,GHz radio continuum. The offsets are with respect to Sgr\,A*.
These absorption maps centered at $+$46 and $+$25 \kms\ illustrate the extent of
the $+$20 and $+$50 \kms\ clouds better than \citet{karlsson03}.  No
absorption is observed toward Sgr\,A* or the inner part of the CND
(the minispiral). See \citet{pihlstrom06} for our absorption results
over the full velocity range of the CND. \label{absorption}}
\end{figure*}

The large scale absorption distribution illustrates conclusion
\emph{i)} above. Figure \ref{absorption} plots selected velocity maps
of the 1720\,MHz OH absorption superimposed on the radio continuum
where each panel displays spectrally smoothed absorption with a width
of 21.7 \kms. It resembles the on-line material by K03 (their Fig.~4),
but is included here as it shows the large extent of the 1720\,MHz
absorption more clearly than in K03.  Figure~\ref{absorption}
demonstrates that the large scale absorption at $+$58 to $+$36 \kms\
occurs in the eastern part of the SNR shell, while absorption at lower
velocities ($+$36 to $+$14 \kms) also covers the western part of the
shell, consistent with \emph{i)}. Further, the K03 data as well as our
absorption data (not shown here; see \citealt{pihlstrom06}) reveal the
CND in absorption at high absolute velocities ($\sim$ 100--150 \kms)
at opposing azimuth angles of the CND major axis, thus concluding
\emph{ii)}.  Finally, our data confirm that the minispiral and Sgr\,A*
are devoid of absorption. We refer to K03 (their Sect.~3.1.1) for
discussing the lack of absorption in detail, but note that our results
support that the general lack of absorption toward Sgr\,A* and the
minispiral therefore imply \emph{iii)}.

\section{Discussion}\label{discussion}

The discussion concentrates on the nature of the 1720\,MHz maser
emission, detected at multiple spatial locations in the GC
region. However, the absorption measurements play an important role in
untangling the 3D structure of the region and the origin of the
different groups of masers observed.

\subsection{Structure of the GC region}

Our absorption measurements corroborate the absorption results and
conclusions of K03: parts of the molecular clouds must be {\em
between} the CND in the front, and the SNR in the back
(Sect.~\ref{resabs}, Sect.~\ref{cndmasers}). Previously, the location
of the components in the line-of-sight toward the Sgr\,A Complex have
been subject to different interpretations, in particular the
line-of-sight location of the CND, the molecular cloud material and
the Sgr\,A\,East SNR. Many authors \citep[e.g.,][]{gusten81, ho85,
zylka90, ho91, marshall95, coil00, mcgary01, karlsson03, vollmer03,
christopher05, herrnstein05, tsuboi06} have tried to picture a
three-dimensional (3D) model of the different components and
structures. The latest is presented by \citet{lee08}, who prefer a
model where the Sgr\,A\,East SNR directly pushes against the CND. A
slab of compressed molecular (or atomic) material might be a part of
this interaction, but would not necessarily separate the SNR from the
CND physically.  This model is mostly based on larger velocity
dispersions and lower velocity centroids in selected H$_2$ emission
slit observations, as compared to what is observed for the NH$_3$
density clumps \citep[e.g.,][]{ho91,coil00,herrnstein05}. However,
this does not directly imply that the SNR pushes the CND toward the
observer, as indicated by \citet{lee08}. It does indeed mean that the
H$_2$ gas is hotter, but it might as well be part of a different
kinematic structure.

Since \citet{sandqvist70,sandqvist74}, it is clear that with the many
observations and interpretations of the Sgr\,A\,East Complex using
different tracers and methods over the past $\sim$40 years, a full
review with perhaps more sensitive (re)observations and/or minute
re-interpretations of the available data and facts is needed to
explain the structure. That is beyond the scope of this paper, as here
we just want to explain the nature of the 1720\,MHz OH masers,
but we note that the full 3D structure still is not completely
clear. A 3D model should also be based on information from absorption
lines which yields direct information of the line-of-sight location of
the absorbing gas with respect to the background continuum. We
therefore endorse the result of K03, that the Sgr\,A\,East SNR and the
CND do not necessarily physically interact, since part of the
molecular material lies between them.

\subsection{Origin of the 1720\,MHz masers in the Sgr\,A Complex}

Maser emission is primarily constrained by the narrow range of
physical parameters required (Sect.~\ref{pump}).  The amplification
and population inversion of the 1720\,MHz OH line requires number
densities of the order of $n_{\rm OH} \sim$ 10$^5$~cm$^{-3}$,
temperatures in the range $T$ = 50--125 K and OH column densities of
$N_{\rm OH} \sim$ 10$^{16}$--10$^{17}$~cm$^{-2}$ \citep{elitzur76,
gray91, gray92, pavlakis96, lockett99, pihlstrom08}. Such conditions
can be found in SFRs \citep{palmer84, baudry88, gray91,gray92,
cohen95, niezurawska04} and in the post-shock regions near SNRs
\citep{frail94,green97,wardle99}.

Modeling of SNR/ISM interactions show that C-type shocks provide the
required inverted OH column densities. However, in this paper we argue
that some masers arise in non-standard locations, i.e., in the CND
(Sect.~\ref{cndmasers}). An independent check on whether this is
reasonable can be made by estimating the OH column density in regions
where masers occur, using the absorption data. In local thermal
equilibrium (LTE)
\begin{equation} 
N_{\rm OH}=1.96\times10^{15}\,T_{\rm ex}\,\Delta V_{\rm FWHM}\,\tau_{\rm peak}~{\rm cm}^{-2} 
\end{equation}
where $T_{\rm ex}$ is the excitation temperature in K, $\Delta V_{\rm
FWHM}$ is the FWHM line width in \kms\ and $\tau_{\rm peak}$ is the
1720\,MHz peak optical depth. The opacity is dependent on the
absorbed flux density as well as the flux density of the background
continuum. Since it is difficult to properly separate the different
continuum components (SNR shell, thermal emission from the ionized gas
and non-thermal emission from Sgr\,A*) and their location relative to
the absorbing gas, the first order estimates given below are apparent
opacities given for the assumed, simplest case with all absorbing gas
in front of the continuum (see K03 for $-T_L/T_C$ maps). Another
caveat is, of course, that the OH is not likely to obey LTE conditions
in these regions.

To estimate a typical column density toward the SNR shell, a
(northeastern) region centered on $\Delta$RA = 75\arcsec\ and
$\Delta$Dec = 104\arcsec\ was chosen. Here, the absorption feature has
a velocity centroid close to $+$50 \kms\, thus corresponding to gas in
which the SNR/ISM masers arise. A single-component Gaussian fit yields
an opacity of 0.7$\pm$0.2, $\Delta V_{\rm FWHM} =$ 15$\pm$3 \kms\ and
an OH column density $N_{\rm OH} =$
2.2$\pm$0.7$\times$10$^{16}\,T_{\rm ex}$~cm$^{-2}$, which are typical
values in other regions toward the Sgr\,A\,East shell.  Similarly, the
OH column density of the CND is estimated at a position centered at
$\Delta$RA = $-$18\arcsec\ and $\Delta$Dec = 8\arcsec. To ensure the
CND opacity is measured excluding absorption from the molecular belt,
a line with a centroid velocity of 98 \kms\ was used, resulting in
$\Delta V_{\rm FWHM} =$ 31$\pm$3 \kms, an opacity of 0.5$\pm$0.1 and
an OH column density of $N_{\rm OH} =$
2.7$\pm$0.7$\times$10$^{16}\,T_{\rm ex}$~cm$^{-2}$.  Compared to the
SNR/ISM interaction sites, the line widths appear broader in the CND,
or alternatively consist of a larger number of overlapping components
in velocity. Overall, we deduce that the estimated OH column densities
in the CND and in the SNR/ISM are similar, providing similar column
density conditions.

The expected temperatures of 50-125 K can be combined with our
estimate of the column density in the CND, $N_{\rm OH}\simeq$
2$\times$10$^{16}\,T_{\rm ex}$~cm$^{-2}$ (Sect.~\ref{resabs}). The
resulting observed column density is an order of magnitude larger than
the $N_{\rm OH} \sim$ 10$^{16}$--10$^{17}$~cm$^{-2}$ predicted by
1720\,MHz maser theories for regions with number densities of
10$^5$~cm$^{-3}$.  However, this is not necessarily a contradiction,
since the masers only occur in regions where the number density is
constrained to $n_{\rm OH} \sim$ 10$^5$~cm$^{-3}$. In regions with
number densities above this value (reflected in an increased column
density) maser emission will be quenched, perhaps in favor of
1612\,MHz emission \citep{pihlstrom08}, and the gas will be observed
in absorption instead. We note that \citet{pihlstrom08} do comment on
1612\,MHz emission toward Sgr\,A\,East, but none is detected toward
the CND.

\subsection{The SNR masers}\label{snrmasers}

This subsection discusses the SNR masers found in the $+$30 to $+$70
\kms\ velocity range; first the new and previously known masers in the
northwest (crosses in Fig.~\ref{maserpos}), then the newly found
masers in the northeast (squares), and finally the previously known
masers in the southeast (plusses).

\subsubsection{Northwestern SNR masers}\label{snrnw}

\begin{figure*} \begin{center}
\plotone{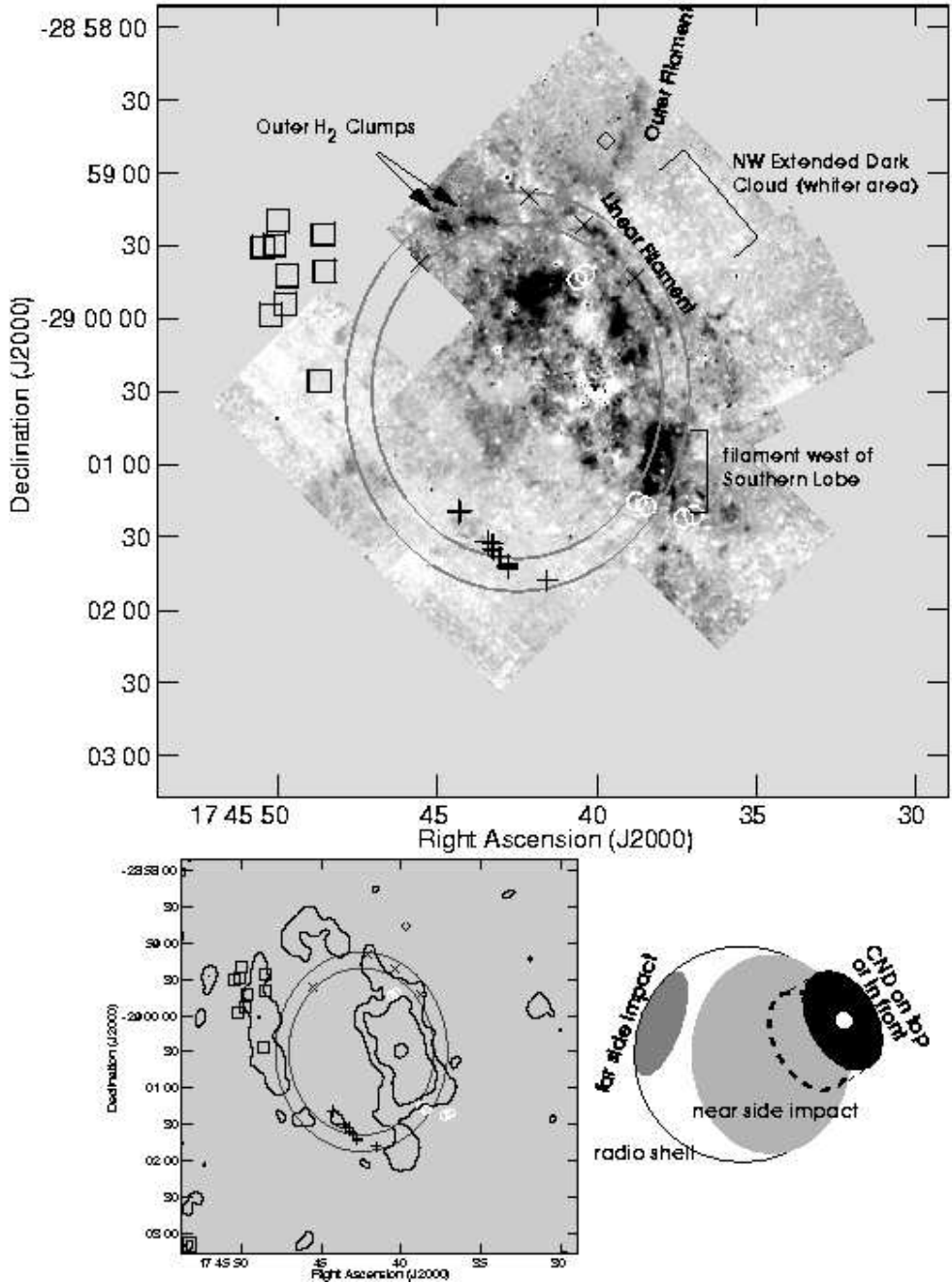} 
\end{center}
\caption{\emph{Top:} Indicators of shock-excited material -- 1720\,MHz
maser positions overlaid on the H$_2$ map of
\citet{yusef-zadeh01}. Symbols as in Fig.~\ref{maserpos}. The
northwestern masers (crosses) can suggestively be fitted with a
(partial) band (between the gray ellipses) that includes the near side
shock excited regions north and west of the CND traced by the H$_2$
filaments described by \citet{yusef-zadeh01} (their Fig.~2b; see
text). The southeastern masers (plusses) suggestively would also fit
if the SNR G359.02$-$0.09 (Fig.~\ref{maserpos}) would not have pushed
them inward, toward the northwest.  \emph{Lower left:} A redisplay of
Fig.~\ref{maserpos} as a contour map without labels but with the
elliptical band for comparison.  Note that the radio continuum shell
is larger than the elliptical band; the radio continuum covers the
northeastern masers (squares) at the far side impact.  \emph{Lower
right:} A schematic model of Sgr\,A\,East interacting at the near side
(light gray filled ellipse) and at the far side (dark gray filled
ellipse). The black filled ellipse and the white dot depict the CND
and Sgr A*. Whether the elliptical band locating the region of
interaction on the near side is complete or not, it depicts that the
near side interaction region is larger than the size of the CND. We
cannot determine whether the CND touches the near side impact region
or not, only that it is located on its front side.  The Southern Lobe
of the CND overlaps with the elliptical band confusing the picture; it
incorrectly suggests that the two CND masers (white circles) in
projection seen in the band are also due to the interaction.
\label{ellipse} }
\end{figure*}

The masers to the northwest, north of the CND, appear related to
linear filaments traced by H$_2$ emission (usually presumed to trace
shocked, dense gas with $n\sim 10^5$~cm$^{-3}$). The positional
coincidence of masers and H$_2$ filaments was first noted by
\citeauthor{yusef-zadeh01} (2001, hereafter YZ01), who used NICMOS
observations to compare the position of the $+$42 \kms maser (\#5,
YZ96) with the distribution of H$_2$ line emission. The maser was
found co-located with an H$_2$ filament labeled the Linear
Filament. In Fig.~\ref{ellipse} we overplot the same H$_2$ map with
all masers. We note that the $+$32 \kms\ maser (\#8, K03) is projected
along the same Linear Filament as maser \#5, and that the new $+$100
\kms\ maser (\#6) is located just north of another filament, called
the Outer Filament.  The position and velocity of the $+$100 \kms\
maser may indicate that it is related to the $+$70 \kms\ cloud
\citep{gatley86} or to the NW Extended Dark Cloud, but no H$_2$
velocity information is available for this cloud for comparison
(YZ01).

YZ01 suggest that the H$_2$ filaments could be generated by the impact
of Sgr\,A\,East into the backside of the CND. As a note of caution,
they also point out that this region is confused by the presence of
the $+$70 \kms\ cloud and the NW Extended Dark Cloud, either of which
may be falling into the CND from the foreground
(\citealt{gatley86,jackson93,marshall95}; YZ01; K03). YZ01 base their
interpretation on a spatial alignment of the Linear Filament with the
western boundary of the SNR radio continuum ridge. A similar alignment
can, however, also be argued with the NW Extended Dark Cloud.  Thus,
the suggestion that the H$_2$ filaments are produced by a direct
impact between Sgr\,A\,East and the CND seems viable but, at this
point, lacks unambiguous evidence.

To explain the appearance of the H$_2$ filaments across the whole
eastern part (including the Linear Filament, the Outer H$_2$ clumps,
and the filament west of the Southern Lobe) in connection to the
masers in this region, we postulate an alternative model where the SNR
triggers a shock in something much larger than the CND. This model is
depicted in the lower right inset of Fig.~\ref{ellipse}. Such an
arc-like interaction model (outlined by the elliptical band in
Fig.~\ref{ellipse}) would argue in favor of an impact of Sgr\,A\,East
into more widespread molecular material, located between the SNR and
the CND. This interaction could, but would not necessarily have to
shock the CND.

As mentioned before, line emission observations (whether from H$_2$ or
NH$_3$ gas, or OH masers) alone do not allow a proper line-of-sight
determination of the different components --- we are looking forward
to e.g.\ absorption and/or extinction measurements toward these H$_2$
filaments and H$_2$ clumps. For now, we can state that the H$_2$
emission filaments are consistent with \emph{any} of these
interpretations: \emph{a)} Sgr\,A\,East is interacting with the CND,
\emph{b)} Sgr\,A\,East is interacting with the $+$20/50 \kms\ cloud in
between the SNR and the CND, or \emph{c)} the H$_2$ filaments are not
due to an interaction with Sgr\,A\,East but due to an interaction with
the NW Extended Dark Cloud or the $+$70 \kms\ cloud. Nevertheless,
with the absorption data available, and a model as shown in
Fig.~\ref{ellipse} (lower right panel), we favor alternative \emph{b)}
where these northwestern SNR masers are generated in the molecular
cloud material at the front (near) side of the SNR. We cannot
determine whether the shock does, or does not penetrate the CND from
the back.

\subsubsection{Northeastern SNR masers}\label{snrne}

One maser excepted, the group of weaker masers in the northeastern
part of Sgr\,A\,East are all new detections. These masers appear
situated near the well known ultra compact \ion{H}{2} regions located
farther east of the SNR (A, B, C, \& D; \citealt{ho85}). Therefore,
these northeastern masers could be associated with star formation,
similar to what is observed in SFRs \citep{niezurawska04,
szymczak04}. Relative to the SNR however, the \ion{H}{2} regions are
in the foreground (K03). Furthermore, the spatial spread and velocity
distribution of the masers follow the ``backward C'' shape observed in
NH$_3$ by \citet{ho91} with the most red-shifted maser velocities in
the center of the distribution. This combination thus depicts the
shell-like interaction of the back\footnote{If the most blue-shifted
maser(s) would be in the center and the NH$_3$ velocity signature
would have a regular C shape \citep{ho91}, this would have been the
front side, propagating toward the observer.}  side of the
Sgr\,A\,East SNR plowing into the $+$50 \kms\ cloud and propagating
away from the observer. This interaction and morphology has been
described previously by many authors and indeed also implies that part
of the $+$50 \kms\ cloud is behind Sgr\,A\,East
\citep[e.g.][]{whiteoak74,sandqvist89,zylka90,ho91,lee08}.  This part
of the $+$50 \kms\ cloud behind the SNR has also been referred to as
the (eastern part of the) Sgr\,A\,East core \citep{zylka90,mezger96}.
The detection of the 1720\,MHz OH masers independently confirms that
there is such an interaction region and that the excitation mechanism
for the observed H$_2$ emission is indeed due to a C-type shocked
front \citep{lee03,lee08}.  Unfortunately, H$_2$ filaments cannot be
recognized in their slit observations (Sect.~\ref{snrnw}). In summary,
we conclude that the northeastern SNR masers are generated in the
molecular cloud material at the back (far) side of the SNR.

\subsubsection{Southeastern SNR masers}\label{snrse}

The previously reported bright SNR masers were all detected in our
observations (YZ96; YZ01; K03). Those are the bright masers near the
southeastern edge of Sgr\,A\,East (Fig.~\ref{maserpos}). The
near-linear distribution of the masers in the southeast differs from
the more spread-out spatial distribution of the masers to the
northeast (Sect.~\ref{snrne}).  In the southeast corner of
Fig.~\ref{maserpos}, an $\sim$80\arcsec-radius semicircular weak
continuum feature can be discerned.  This continuum emission outlines
another SNR shell, G359.02$-$0.09 \citep[e.g.,][]{coil00,
herrnstein05}. \citet{sakano03} have detected non-thermal X-ray
emission from parts of the G359.02$-$0.09 shell, with a spectrum that
suffers large absorption consequently placing the X-ray source in the
GC. In the sky, the G359.02$-$0.9 SNR shell overlaps the Sgr\,A\,East
continuum, and is likely responsible for the inward-concave morphology
of the southeastern Sgr\,A\,East shell \citep{coil00}. This is
precisely where the bright 1720\,MHz masers form. In contrast to the
northwestern masers, the NICMOS image only shows weak extended H$_2$
emission, no filaments (Fig.~\ref{ellipse}; YZ01). Moreover, this
region displays almost no continuum background and OH absorption
(Fig.~\ref{absorption}; Fig.~\ref{ellipse}; K03). This region likely
is a result of two colliding shock fronts and would explain why there
is such a sharp line of bright masers. Previous discussions of these
masers suggested that they are the result of the interaction of the
$+$50 \kms\ cloud and Sgr\,A\,East. We here suggest that they probably
are due to a SNR/SNR interaction (albeit close to the front-side
SNR/ISM interaction in Fig.~\ref{ellipse}), for which 1720\,MHz maser
emission has not been anticipated before.

\subsection{The CND masers}\label{cndmasers}

We here discuss the masers near the tangent points of the CND with
high absolute ($\pm$104--141 \kms) velocities (circles in
Fig.~\ref{maserpos}).

\subsubsection{Origin and variability of the CND masers}

A bright maser at $+$132 \kms\ has been detected in all observations
(i.e., since 1986 and beyond 2005) in the northern part of the disk
(YZ96; K03; this work). Previously, only in 1986 a faint, conjugate $\
-$132 \kms\ maser was detected in the southern part of the CND (K03
and references therein) which disappeared before the next possible
detection in 1996. Up until now, this single detection has been the
only observation suggesting that some 1720\,MHz masers are distributed
symmetrically in position and velocity with respect to Sgr\,A*, like
the CND, in sharp contrast to the SNR/ISM interaction at velocities
between 30 and 70 \kms\ as discussed above. The three new detections
of high-negative velocity 1720\,MHz masers toward the southwestern part of the
CND (Table~\ref{masers}) now strongly support the existence of
conjugate masers with locations and velocities consistent with
originating from gas in the CND. Clearly these masers are variable as
they may appear and/or disappear on time scales on the order of years
(Sect.~\ref{resmas}).

The estimated OH column density derived above appears similar across
the total extent of the CND, indicating that maser emission
potentially could be observed at all azimuthal angles of the CND. This
in particular would be the case if Sgr\,A\,East would continuously
drive planar shocks from behind into the CND while the CND is pushed
toward the observer as argued by YZ01 and \citet{lee08}\footnote{Note
that the \citet{lee08} H$_2$ data do not properly cover the CND.}.  In
such a case, one would expect to find some masers originating in the
$+$20 \kms\ cloud, i.e., with velocities in the $\sim$ 10--50 \kms\
range, in projection \emph{toward the whole CND}. Though there are
three masers in this velocity range north of the CND (and associated
with H$_2$ filaments), there is no distribution toward the rest of the
CND. Furthermore, a push from behind would result in a relatively
constant pumping of the masers, which due to the clear variability of
the southwestern (high-negative velocity) CND masers is hard to
support. In contrast, the OH masers in the CND appear only in two
regions: slightly east of north and west of south at positions close
to the geometrical tangent points. Extended H$_2$ emission is found
nearby those regions, but not distributed in filaments as for the
northwestern masers (Sect.~\ref{snrnw}). This H$_2$ emission is due to
both limb brightening effects as well as dissipation of shocks in
clump-clump interactions (YZ01). For the case of masers in a disk or
ring structure, this geometry can be understood by the requirement of
having long paths of velocity coherent gas in order to build up a
large amplification. Closer to the tangent points of a disk, the
path-lengths are longer resulting in a larger amplification
\citep{parra05}. Such a behavior has previously been observed in maser
transitions of H$_2$O and OH in extragalactic sources
\citep{miyoshi95, pihlstrom01, kloeckner03}. It is therefore plausible
that the CND masers can be sustained without being a direct result
from shocks generated by Sgr\,A\,East (Sect.~\ref{pump}).

\subsubsection{Pump source of the CND masers}\label{pump}

Whereas the pumping of the SNR masers can be explained by the
interaction of the SNR shock plowing into the interstellar material of
the $+$50 \kms\ cloud, it is not required that the SNR and the CND
directly interact to form the CND masers.

A radiative (far-infrared) pump source primarily tends to invert the
1612\,MHz line, while collisional excitation has been identified as
the prime pumping mechanism for 1720\,MHz OH masers
\citep{elitzur76}. The lack of interstellar OH 1612\,MHz and mainline
emission in the CND (K03) argues against far-infrared pumping and for
a collisional pumping scheme of the 1720\,MHz CND masers. In the CND
we find extended shock excited H$_2$ emission in the ``Lobes'' (YZ01),
but not in large-scale shock-front filamentary structure as for the
northwestern masers. The clumpiness and irregular distribution of
molecular gas in the CND imply the possibility of frequent clump-clump
collisions which could provide the source of a collisional maser
pump. The dense ($n_{\rm H_2}\simeq$ 3--4$\times$10$^7$~cm$^{-3}$)
molecular clumps defined by HCN and HCO$^+$ observations are found
more or less in the same locations where bright H$_2$ emission has
been mapped (\citealt{wright01,christopher05,jackson93}; YZ01).  YZ01
use the H$_2$ line emission associated with the CND to argue that
dissipation of the random motions of molecular clumps is the most
likely cause of the excitation of the H$_2$ molecules via shocks,
implying the presence of C-type shocks within the CND. Similar to the
C-shock chemistry predicted for SNR/ISM masers, the post-shock regions
in the CND should produce suitable conditions for regions of enhanced
OH abundances \citep{wardle99}.

Comparing the maser positions with respect to the H$_2$ (YZ01) and HCN
emission \citep{christopher05} and assuming that the CND masers are
excited via shocks, it is not surprising to see a correlation between
the maser emission and the H$_2$ emission. It is interesting to note
that the masers always appear to be offset from, and trailing the
brightest HCN and H$_2$ peaks. At the high densities ($n_{\rm H_2}
\sim 10^7$~cm$^{-3}$) traced by HCN, maser emission will be
quenched. Instead, the masers occur in lower density post-shock
regions following/trailing the clumps. The large dispersion ($\sim$ 40
\kms) in the southwestern CND masers may reflect the dispersion of
individual clumps in the CND.

A clumpy medium also appears to be a better maser amplifier than a
smooth medium. Extragalactic OH masers have successfully been modeled
assuming a clumpy medium \citep{pihlstrom01, parra05}. Modeling of
masers in a clumpy medium predicts high maser variability caused by
smaller regions of inverted gas moving in and out of the
line-of-sight. However, no clear limits could be set on the
variability of individual maser features presented in this paper as
each observation so far has very different observing parameters
(Table~\ref{datasets}).  A forthcoming paper will properly address the
maser variability using homogeneous data sets.

\section{Summary}

The velocities and locations of most 1720\,MHz OH masers observed in
the GC agree with the commonly adopted model in which the masers arise
in regions where the supernova remnant Sgr\,A\,East is interacting
with the $+$20 and $+$50 \kms\ molecular clouds and the nearby SNR
G359.02$-$0.09. In addition to extending the database of such masers,
in particular toward the interaction region toward the northeast rim
of Sgr\,A\,East, we have explored the slightly different possible
origins of the maser groups, notably the masers northwest of the
CND. We showed that they indicate a region of shock interaction of the
near side of Sgr\,A\,East with molecular cloud material extending well
over the projected size of the CND. The newly found masers in the
northeast indicate such an interaction on the far side of
Sgr\,A\,East.

Furthermore, we have verified the existence of high-negative velocity
masers, which together with previously detected high-positive velocity
masers bracket Sgr\,A* both in position and velocity. Their positions
and velocities are consistent with being located near the tangent
points of the CND. However, we find no need for the SNR to interact
directly with the CND to pump and sustain these high-velocity
masers. OH absorption data show that the CND and Sgr\,A\,East are
separated by molecular cloud material, and therefore the SNR shock is
unlikely to be responsible for pumping these high-velocity
masers. Instead a more likely pumping mechanism is by collisions from
dissipation between clumps \emph{in} the CND.

These results imply that 1720\,MHz OH masers \emph{not only} do occur
in SFRs and in SNR/ISM (and SNR/SNR) interactions, but can also arise
in clumpy and disturbed circumnuclear disks at small radii. If
sufficiently bright, such 1720\,MHz masers could serve as an
additional molecular line tool to study circumnuclear gas dynamics
around galaxy centers.

\section*{Acknowledgments} 

We would like to thank M.~Elitzur and V.~Fish for their most useful
comments on an earlier version of the manuscript, and also S.~Stolovy
and M.~Christopher for kindly sharing their H$_2$ and HCN maps for
reference. The Very Large Array is operated by the National Radio
Astronomy Observatory, which is a facility of the National Science
Foundation operated under cooperative agreement by Associated
Universities, Inc.

{\it Facility:} {VLA}

 \end{document}